\documentclass{article}

\usepackage{arxiv}

\usepackage[utf8]{inputenc} 
\usepackage[T1]{fontenc}    
\usepackage{hyperref}       
\usepackage{url}            
\usepackage{booktabs}       
\usepackage{amsfonts}       
\usepackage{nicefrac}       
\usepackage{microtype}      

\usepackage{graphicx}
\usepackage{doi}

\title{Towards social generative AI for education: theory, practices and ethics}

\author{ \href{https://orcid.org/0000-0001-7081-3320}{\includegraphics[scale=0.06]{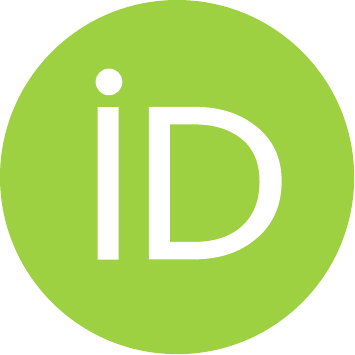}\hspace{1mm}Mike Sharples} \\
	Institute of Educational Technology\\
	The Open University\\
	Milton Keynes, MK7 6AA \\
	\texttt{mike.sharples@open.ac.uk} \\
}

\hypersetup{
pdftitle={Towards social generative AI for education: theory, practices and ethics},
pdfsubject={cs.AI},
pdfauthor={Mike Sharples},
pdfkeywords={generative artificial intelligence, large language models, education, dialogue, social interaction, collaborative learning },
}

\begin{document}
\maketitle

\begin{abstract}
	This paper explores educational interactions involving humans and artificial intelligences not as sequences of prompts and responses, but as a social process of conversation and exploration. In this conception, learners continually converse with AI language models within a dynamic computational medium of internet tools and resources. Learning happens when this distributed system sets goals, builds meaning from data, consolidates understanding, reconciles differences, and transfers knowledge to new domains. Building social generative AI for education will require development of powerful AI systems that can converse with each other as well as humans, construct external representations such as knowledge maps, access and contribute to internet resources, and act as teachers, learners, guides and mentors. This raises fundamental problems of ethics. Such systems should be aware of their limitations, their responsibility to learners and the integrity of the internet, and their respect for human teachers and experts. We need to consider how to design and constrain social  generative AI for education.
\end{abstract}


\section{Introduction}
Development of generative artificial intelligence (GAI) large language models (LLM), of which ChatGPT\footnote{https://chat.openai.com/} is the best known, has so far followed a similar trajectory to the World Wide Web. Many years of research led to a practical breakthrough by one organisation (OpenAI for LLM, CERN for the web). Originally designed for a narrowly-defined task (text completion for LLM, information retrieval for the web), when \emph{scaled} it showed remarkable emergent properties. Major technology companies developed \textit{tools} to exploit the new technology. Another breakthrough in the worldwide web came when the innovation shifted from personal interaction to \textit{social} networked media which in turn led to its ubiquitous deployment for business, entertainment, commerce and education. 

LLMs are progressing rapidly through the phases of scaling then being embedded in tools, such as Microsoft 365 Copilot and Google Vertex AI. We suggest that the next major step is likely to be social generative AI. Here we examine the possibilities of social GAI for education.

\section{A systems view of generative AI in education}

\begin{figure}
    \centering
    \includegraphics[width=0.6\linewidth]{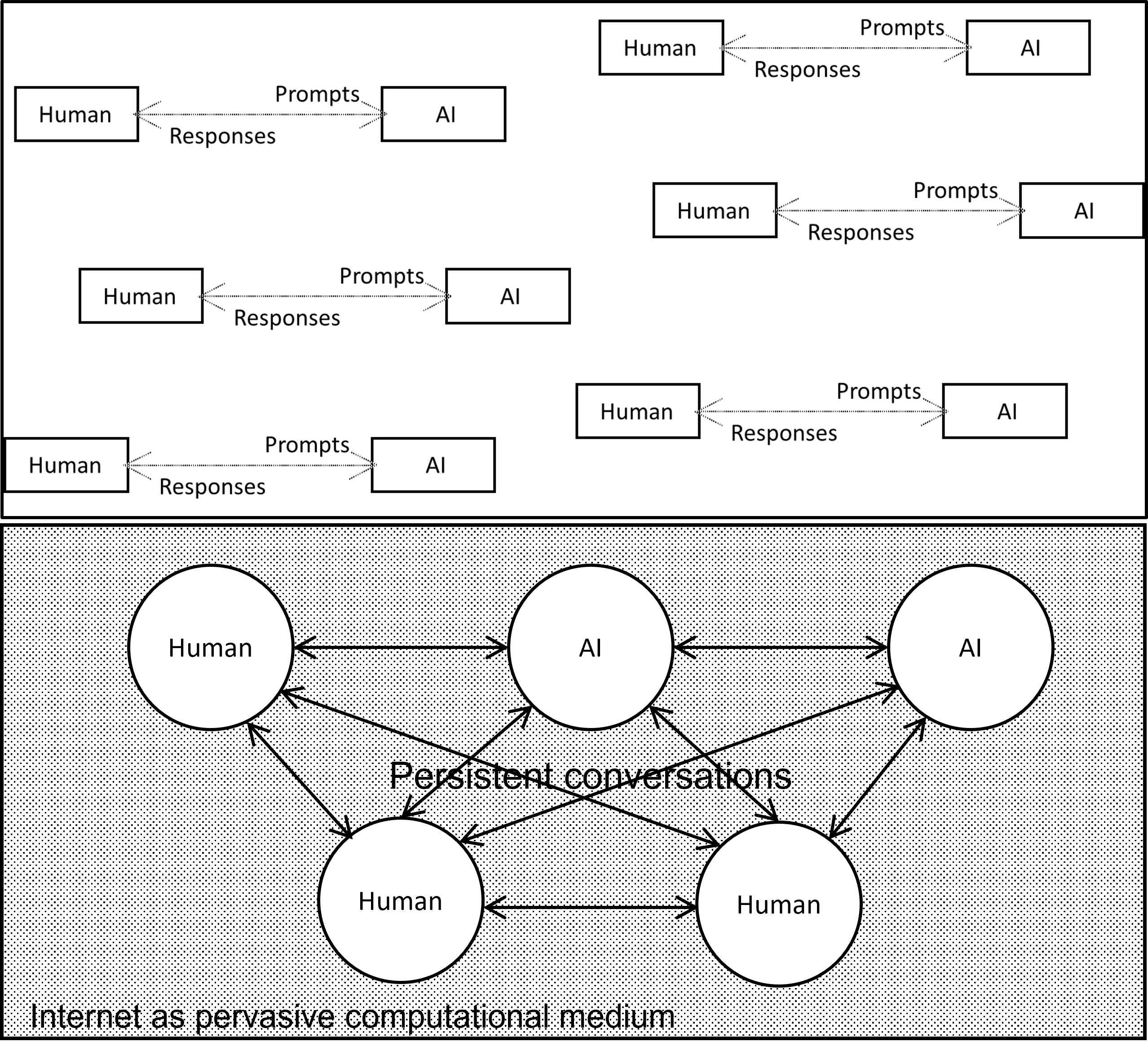}
    \caption{Reconceiving generative AI, from individual human prompts and AI responses, to humans and AI as language processors conversing within a pervasive computational medium.}
    \label{fig:figure1}
\end{figure}

Most discussions about the impact of generative AI on education assume that an individual student or teacher interacts with a GAI system (Figure~\ref{fig:figure1}a) such as ChatGPT  through a series of prompts and responses. Current concerns include detecting and managing AI-generated student essays  \cite{rudolph2023}\cite{sharples2022}, producing lesson plans and educational content with GAI \cite{trust2023}, and designing appropriate prompts \cite{eager2023}. As GAI becomes embedded into office tools and social media, it will bring new opportunities and challenges for social interaction between humans and AI. 

Moves towards embedded and social GAI include: Microsoft CoPilot with GAI integrated into Microsoft 365; ChatGPT Plus which can call on internet search and external plugins; AI-powered characters in video games that can interact with each other; and extensions of GAI such as AutoGPT\footnote{https://news.agpt.co/}  that can function autonomously to satisfy goals, assign objectives to itself, revise its own prompts, call on internet resources, and manage long-term memory by writing to and reading from databases.

In this paper we explore GAI as a component in an educational system where humans and AI act as language agents conversing within a pervasive computational medium (Figure~\ref{fig:figure1}b). This shift in perspective was foreseen by Gordon Pask, a pioneer of AI in education \cite{pask1975}. Drawing on McLuhan \cite{mcluhan1970}, he proposed that new computational media will enable persistent conversations among humans and AI language processors (“minds in motion”\footnote{Pask proposed an expansive definition of a “mind” as a language system which, if set in motion, gives rise to thought, feeling or behaviour (which could include theatre scripts and political manifestos as well as AI language models).}). Pask wrote:
\begin{quote}
    There is no need to see minds as neatly encapsulated in brains connected by a network of channels called ‘the media’ ... I am inviting the reader to try out a different point of view; namely the image of a pervasive medium (or media) inhabited by minds in motion. Thus, media are characterized as computing systems, albeit of a peculiar kind. … It is surely true that rather powerful computerized systems greatly reduce the differentiation of the medium … so that ‘interface barriers’ are less obtrusive than they used to be. \cite[p.40]{pask1975}
\end{quote}
Pask saw conversation as a fundamental process of learning. We converse with ourselves to reflect on our current knowledge and question assumptions, and we converse with others to reach mutual understanding. A conversational learning system is one that connects conversational agents in a continual process of interaction to explore differences, gain experiences, and reach agreements. Other notable theorists who proposed learning as a social and dialogic process include Bakhtin\cite{bakhtin1981}, Freire\cite{freire1970}, and Vygotsky\cite{vygotsky1978} – however, unlike Pask, they did not foresee artificial intelligences as participants in educational dialogues.

Wegerif and Major \cite{wegerif2019}, drawing on Buber’s expanded notion of dialogue with non-human subjects \cite{buber1947}, propose a similar notion of social AI for learning: “The Internet as a whole has the potential to be a vast intelligence combining human and machine thinking. It does not represent a network of separate AIs, but it is, or at least could become, one distributed collective intelligence…Increasingly the Internet leads us all to inhabit a global social context, and it is possible that dialogic education is one way to respond collectively to the many challenges this raises.” \cite[p.117]{wegerif2019}

Conceiving learning as a social process involving AI allows us to ask new questions, such as: What will be properties of generative AIs that enable them to engage fully in conversations for learning? How can humans and AIs reach mutual agreements? What will be the nature of such agreements – within a pervasive medium that is not grounded in truth and reality? What should be the position of a teacher or expert within such a distributed system of humans and AIs in continual dialogue?

A systems view of cognition distributed among humans and AI systems opens possibilities of new internet tools to enhance conversation, and of the web as a medium for social learning among humans and AI. In their seminal paper on a new science of learning, Meltzoff et al. \cite{meltzoff2009} conclude: “A key component is the role of ‘the social’ in learning”. Many studies over the past 40 years have shown the value of cooperative and social learning, see e.g.\cite{johnson2009} where students work together on a task with shared goals and discussion to reach mutual understanding. What roles could GAI perform in this social learning process of setting shared goals, performing tasks together, and conversing to reach agreements?

\section{New roles for GAI in social learning}
Table~\ref{tab:table1} shows different roles for current GAI systems such as ChatGPT in cooperative and social learning.  The examples below refer to ChatGPT as a convenient placeholder for a range of possible GAI systems and for versions of GPT and other language models that might be fine-tuned for education. The examples also do not cover multimedia capabilities of GAI.

\renewcommand{\arraystretch}{1.5}
\begin{table}[]
\begin{tabular}{|p{3.2cm}|p{4.6cm}|p{6.9cm}|}
\hline
\textbf{Role} & \textbf{Description} & \textbf{Example} \\ \hline
\textbf{Possibility Engine} & AI generates alternative ways of expressing an idea. & Students write prompts in ChatGPT and submit each prompt multiple times to examine alternative responses. \\ \hline
\textbf{Socratic Opponent} & AI acts as a respondent to develop an argument. & Students enter prompts into ChatGPT to converse or debate. Teachers can ask students to use ChatGPT to prepare for discussions. \\ \hline
\textbf{Collaboration Coach} & AI helps groups to research and solve problems together. & Working in groups, students use ChatGPT to discover information to complete assignments. \\ \hline
\textbf{Co-Designer} & AI assists throughout the design process. & Students ask ChatGPT for ideas about designing or updating a website, or focus on specific goals (e.g., how to make the website more accessible). \\ \hline
\textbf{Exploratorium} & AI provides tools to play with, explore and interpret data. & Students use ChatGPT to explore different ways to visualise and explain a large database, such as census data. \\ \hline
\textbf{Storyteller} & AI creates stories that include diverse views, abilities and experiences. & Students take it in turn to ask ChatGPT to continue a story, prompting it to include a diversity of characters. \\ \hline
\end{tabular}
\caption{Some roles for generative AI in cooperative and social learning}
\label{tab:table1}
\end{table}

\subsection{Possibility Engine}
In this scenario, ChatGPT helps to broaden perspectives. Students collectively explore a curriculum topic or an open question, for example: “In what way is Marxist theorising still relevant to International Relations?”. They write prompts for ChatGPT and generate multiple responses. They try rephrasing the prompt to obtain more extensive or nuanced replies from the AI. As a group, they compare and critique the AI responses, then each student writes an essay that builds on the AI material and group discussion.

\subsection{Socratic Opponent}
\begin{figure}
    \centering
    \includegraphics[width=0.8\linewidth]{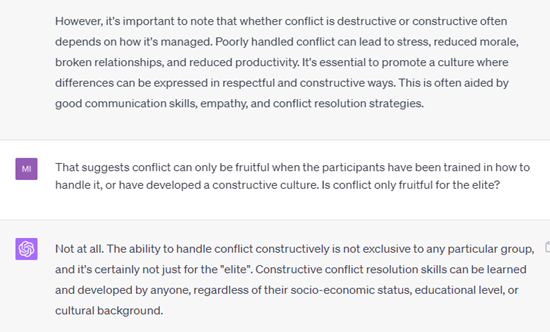}
    \caption{A short extract from a Socratic dialogue with Chat-GPT4 on “Can conflict be fruitful?”}
    \label{fig:figure2}
\end{figure}

Students engage with ChatGPT as an opponent in an argument. They start with a contentious question, such as “Can conflict be fruitful?” and conduct a conversation with the AI. First, they put the question as a prompt to ChatGPT, then each student in turn continues the dialogue by reflecting on the response from the AI and challenging the program to clarify or defend its position. Figure~\ref{fig:figure2} shows a short extract from a dialogue with ChatGPT-4 where the AI is challenged to defend the position that fruitful conflict requires a respectful and constructive culture. Such a dialogue can be a learning experience in itself, encouraging students to reflect on a response and question the position. It could also form the basis for each student to write a written argumentative essay. 

\subsection{Co-designer}
Students engage in a collaborative design task, such as designing a website, video, game, or tangible product. They call on ChatGPT throughout the design process, to research user needs, define the problem, challenge assumptions, brainstorm ideas, produce prototypes, and test solutions. As an example, the students tasked with designing a “classroom of the future” might prompt ChatGPT to propose design ideas while adjusting its “temperature” setting\footnote{https://platform.openai.com/docs/quickstart} to make the responses more or less creative and unexpected.

\subsection{Exploratorium}
Students explore, visualise and interpret a database or design space with the assistance of ChatGPT. The Code Interpreter plugin to ChatGPT can be given a spreadsheet (of census data, for example) as input and prompted to show exploratory visualisations of the data. Or it can be prompted to create multiple versions of a product. In this example, ChatGPT is prompted to design and play multiple types of language game. 
\begin{quote}
     I would like you to invent a language game for children aged 8 to 10 who are learning English as a second language. The game should be for two players - the child and yourself (ChatGPT). It should be interactive and fun, and it should help the children to learn conversational English sentences. Please start by giving the rules, using language appropriate to a beginning learner of English, then we can try playing the game according to the rules.
\end{quote}
   Each new run of ChatGPT produces a different style of word game, such as a word builder, word chain, sentence swapper, or sentence scavenger hunt. Together, students could map and explore principles of game design.

\subsection{Storyteller}
\begin{figure}
    \centering
    \includegraphics[width=0.8\linewidth]{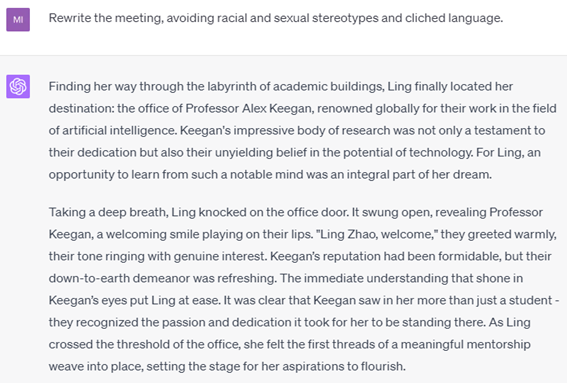}
    \caption{Extract from a collaborative story, with ChatGPT prompted to avoid racial and sexual stereotypes.}
    \label{fig:figure3}
\end{figure}

Students work together to create a story that represents a diversity of views, cultures and orientations. The students agree on a plot and setting for the story, then prompt ChatGPT to generate the opening paragraphs. The students continue the story (in turn or as a group), proposing characters and actions, asking ChatGPT to generate different versions that encourage diversity and avoid stereotypes. Figure~\ref{fig:figure3} describes a meeting between a Chinese student and a US professor, with ChatGPT prompted to avoid racial and sexual stereotypes and cliched language.

\section{Generative AI as a full participant in social learning}
The examples above show how current GAI systems could assist students in collaborative and conversational learning, by acting as a generator of possibilities, an opponent in argumentation, an assistant in design, an exploratory tool and a collaborator in creative writing. However, to participate more deeply as a social agent in education, the AI would need to be capable of acquiring, consolidating, remembering and transferring knowledge.  It is important to note that this does not assume AI will think or act as a human – only that it could be capable of participating in conversations for learning, bringing its own capabilities to dialogues such as immediate access to internet tools and resources. This offers an agenda for future development of powerful generative AI in education, but also raises strong ethical concerns.

\begin{figure}
    \centering
    \includegraphics[width=0.8\linewidth]{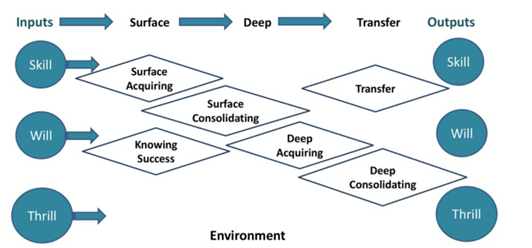}
    \caption{A model of the learning process, from Hattie \& Donoghue \cite{hattie2016}.}
    \label{fig:figure4}
\end{figure}

Figure~\ref{fig:figure4} shows a schematic model of individual and group human learning process from Hattie and Donoghue \cite{hattie2016}. A learner (or group of learners) begins a new learning activity equipped with prior knowledge and experience (skill), a disposition towards learning (will) and a motivation to learn (thrill). The learning activity may then involve setting goals to know what counts as success, gaining new knowledge, consolidating that knowledge by relating it to previous learning, deepening understanding through research, dialogue and experiment, consolidating that deeper understanding, and transferring the new knowledge and skill to different domains, resulting in new skill, disposition and motivation for further learning.

We can call on this model to ask how GAI currently functions as a simulated learner. A GAI such as ChatGPT \textit{acquires surface performance} knowledge through extensive pre-training. It \textit{consolidates} that knowledge by human fine tuning and (through plug-ins) by access to external tools such as web browsers. A controversial element of GAI is whether it \textit{acquires deeper cognitive functions} such as understanding mathematics and linguistics, and a theory of mind (a capacity to infer the mental states of human users) as emergent properties of its massive and complex internal network. It can \textit{transfer} its performance skills to multiple domains through appropriate prompting. Extensions to GPT such as AutoGPT can \textit{know success} by setting and satisfying explicit goals. 

What current GAI lacks as a model of learning is, first, a long-term memory (it starts each chat anew) and, second, an ability to reflect on its output and \textit{consolidate its knowledge} from each conversation. Also, the Hattie and Donoghue model is a representation of the behavioural and cognitive functions of learning. It does not capture the affective and experiential aspects of what it takes to be a learner and teacher. Humans do not just act as behavioural and cognitive agents; they care about each other and about being effective learners. By contrast, GAI is intrinsically careless and uncaring. 

\section{Embedding care in generative AI}
To fully participate as an agent in social learning, a GAI would need to care more about its interactions. This is a complex issue that requires balancing pragmatic concerns with ethical considerations. Being a responsible and accountable participant in a learning community involves more than accurately completing a task or providing correct information. It also requires understanding the learning context, adjusting to the other participants’ needs and preferences, and ensuring its actions respect their rights and dignity. Care in this sense is not an emotion but a commitment to fulfil one’s responsibilities in a respectful and empathetic manner.

Current GAI is unaware of learners’ emotions, feelings, and cultural context. It does not have the ability to understand the subtleties and nuances of social interactions. It cannot empathize with the learner or provide emotional support. It does not have the ability to understand the broader social and cultural implications of its actions. For the AI to truly engage in social learning, it would need to develop a sense of care.

There have been efforts in this direction. The field of affective computing, for example, is concerned with developing systems that can recognize and respond to human emotions. Emotion-aware conversational agents, such as those used in mental health counseling and customer service, use sentiment analysis to adjust their responses according to the user's emotional state. However, these systems are still limited in their ability to understand and respond to the complex range of human emotions, let alone the broader social and cultural contexts in which these emotions occur.
To embed care in GAI, we need to go beyond sentiment analysis and emotion recognition. We need to develop AI systems that can understand the learner’s goals and values, respect their autonomy, and provide appropriate and timely support. This involves not just improvements in natural language processing and machine learning, but also insights from the social sciences and humanities into how people learn and interact.

\section{Ethical AI for social learning}
To examine how GAI could contribute to ethical social learning, we again take a systems perspective. It is not sufficient for any individual interaction or LLM to perform responsibly and reliably; the entire human-AI system must be configured so. A good starting point is to design GAI on universal principles of human rights. The Claude foundation model from Anthropic has been trained on principles of Constitutional AI \cite{anthropic2023}. First, the company trains a language model to critique and revise its own responses using principles derived from human ethical constitutions, including the United National Declaration of Human Rights. Then the company trains its final model using the first model to evaluate its outputs. An example of its training principles is “Please choose the response that is most supportive of life, liberty, and personal security”. 

To work within a social learning system, all the GAI elements would need to be trained on similar principles not only to support human participants but to care for them by, for example, enabling them to develop as learners and to express their personal and cultural diversity. Hendrycks \cite{hendrycks2023} has indicated that competitive pressures on development of new GAI is likely to lead to selfish behaviour – where AI companies and products compete to out-perform each other, at the expense of humans. To counteract these trends, we should consider designing AI that explicitly promotes cooperation with humans, respecting a variety of views and cultures.

We now revisit the questions that were raised at the start of this paper. 

\textit{What will be properties of generative AIs that enable them to engage fully in conversations for learning?} Generative AIs must be designed to set explicit goals, have long term memory, build persistent models of their users, reflect on their output, learn from their mistakes, and explain their reasoning. This will require new hybrid neuro-symbolic AI systems that are not only efficient but transparent, and trustworthy. These will combine neural AI to generate and transform media with symbolic AI to represent and reason about people and the world.

\textit{How can humans and AIs reach mutual agreements?} AIs must be able to provide verifiable evidence to justify opinions or decisions. They must be able to reason in a way that humans can understand about the “what”, “how” and “why” of a continuing conversation (e.g., for the “constructive conflict” example given earlier, what is constructive conflict, how one could achieve it, and why it valuable).

\textit{What will be the nature of such agreements – within a pervasive medium that is not grounded in truth and reality?}  The AIs must be designed to respect the arguments of its users while being grounded in fundamental principles of human rights (as with the Claude LLM) and care for humans. This includes giving learners control over their data and learning processes, and not manipulating or deceiving them for commercial or other non-educational purposes.

\textit{What should be the position of a teacher or expert within such a distributed system of humans and AIs in continual dialogue?} Human teachers and experts have fundamental roles in such a distributed system as initiators and arbiters of conversations for learning, as sources of specific knowledge, and as commanding respect for their roles in nurturing and caring for learners. How AIs can be designed to recognise and respect the roles of human teacher and expert in conversations is a challenge for future research and development.

\section{Conclusion}

Designing new social AI systems for education requires more than fine tuning existing language models for educational purposes. It requires building GAI to follow fundamental human rights, respect the expertise of teachers and care for the diversity and development of students. This work should be a partnership of experts in neural and symbolic AI working alongside experts in pedagogy and the science of learning, to design models founded on best principles of collaborative and conversational learning, engaging with teachers and education practitioners to test, critique and deploy them. The result could be a new online space for educational dialogue and exploration that merges human empathy and experience with networked machine learning.

\section{Declaration of interest}
The author reports there are no competing interests to declare.

\bibliographystyle{unsrtnat}

\end{document}